\documentclass[aps,11pt,prd,groupedaddress,nofootinbib,notitlepage,eqsecnum,preprintnumbers]{revtex4-2}

\usepackage[utf8]{inputenc}
\usepackage{amsmath}
\usepackage{amssymb}
\usepackage{graphicx}
\usepackage{dcolumn}   % needed for some tables
\usepackage{bm}        % for math
\usepackage{amssymb}   % for math
\usepackage{mathtools}
\usepackage{xcolor}
\usepackage{hyperref}
\usepackage{ulem}
\usepackage{comment}

\newcommand{\rh}{\mathrm{rh}}
\newcommand{\Rcal}{\mathcal{R}}
\newcommand{\Pcal}{\mathcal{P}}
\newcommand{\Acal}{\mathcal{A}}
\newcommand{\GW}{\mathrm{GW}}
\newcommand{\PTA}{\mathrm{PTA}}

\begin{document}

\title{Induced Gravitational Wave interpretation of PTA data: \texorpdfstring{\\} a complete study for general equation of state}
    \author{Guillem Domènech${}^{a,b}$}\email{guillem.domenech@itp.uni-hannover.de}
    \author{Shi Pi${}^{c,d,e}$} \email{shi.pi@itp.ac.cn} 
    \author{Ao Wang${}^{c,f}$}\email{wangao@itp.ac.cn}
    \author{Jianing Wang${}^{c,f,g}$}\email{wangjianing@itp.ac.cn}

        \affiliation{
            $^{a}$ Institute for Theoretical Physics, Leibniz University Hannover, Appelstraße 2, 30167 Hannover, Germany.}
        \affiliation{
            $^b$ Max-Planck-Institut für Gravitationsphysik, Albert-Einstein-Institut, 30167 Hannover, Germany}
		\affiliation{
        $^{c}$ CAS Key Laboratory of Theoretical Physics, Institute of Theoretical Physics, Chinese Academy of Sciences, Beijing 100190, China}
        \affiliation{
            $^{d}$ Center for High Energy Physics, Peking University, Beijing 100871, China}
       
                \affiliation{
		$^{e}$ Kavli Institute for the Physics and Mathematics of the Universe (WPI), The University of Tokyo, Kashiwa, Chiba 277-8583, Japan}
 \affiliation{
        $^{f}$ School of Physical Sciences, University of Chinese Academy of Sciences, Beijing 100049, China}
\affiliation{
$^g$ International Center for Quantum-field Measurement Systems for Studies of the Universe and Particles (QUP, WPI),
High Energy Accelerator Research Organization (KEK), Oho 1-1, Tsukuba, Ibaraki 305-0801, Japan}

	\date{\today}

\begin{abstract}
We thoroughly study the induced gravitational wave interpretation of the possible gravitational wave background reported by PTA collaborations, considering the unknown equation of state $w$ of the early universe. We perform a Bayesian analysis of the NANOGrav data using the publicly available \textsc{PTArcade} code together with \textsc{SIGWfast} for the numerical integration of the induced gravitational wave spectrum. We focus on two cases: a monochromatic and a log-normal primordial spectrum of fluctuations. For the log-normal spectrum, we show that, while the results are not very sensitive to $w$ when the GW peak is close to the PTA window, radiation domination is out of the $2\sigma$ contours when only the infra-red power-law tail contributes. For the monochromatic spectrum, the $2\sigma$ bounds yield $0.1\lesssim w\lesssim0.9$ so that radiation domination is close to the central value. We also investigate the primordial black hole (PBH) abundance for both monochromatic and log-normal power spectrum. We show that, in general terms, a larger width and stiffer equation of state alleviates the overproduction of PBHs. No PBH overproduction requires $w\gtrsim0.42$ up to 2-$\sigma$ level for the monochromatic spectrum. Furthermore, including bounds from the cosmic microwave background, we find in general that the mass range of the PBH counterpart is bounded by $10^{-5} M_\odot\lesssim M_{\rm PBH}\lesssim10^{-1} M_\odot$. Lastly, we find that the PTA signal can explain the microlensing events reported by OGLE for $0.42\lesssim w\lesssim 0.50$.   Our work showcases a complete treatment of induced gravitational waves and primordial black holes for general $w$ for future data analysis.
\end{abstract}

\maketitle
\newpage

\section{Introduction}

The world-wide pulsar timing array (PTA) collaborations \cite{EPTA:2023fyk,EPTA:2023sfo,EPTA:2023xxk,Zic:2023gta,Reardon:2023gzh,Reardon:2023zen,NANOGrav:2023hde,NANOGrav:2023gor,InternationalPulsarTimingArray:2023mzf,Xu:2023wog} reported tentative evidence of a low-frequency Gravitational Wave (GW) background. Although an astrophysical GW background is expected from inspiraling supermassive black hole binaries \cite{NANOGrav:2023hfp},\footnote{See though Refs.~\cite{Huang:2023chx,Gouttenoire:2023nzr,Depta:2023qst} interpreting the GW background as the merger of supermassive primordial black holes. } the simplest predictions seem to be in mild tension with the data \cite{NANOGrav:2023hvm}. Other plausible explanations of the GW background have a cosmological origin \cite{Sazhin:1977tq,Caprini:2018mtu,Roshan:2024qnv,Kawai:2023nqs}. For instance, one of the cosmic interpretations that seems to behave better in the Bayesian analysis of NANOGrav \cite{NANOGrav:2023hvm} are GWs induced by primordial curvature fluctuations (see also Refs.~\cite{Dandoy:2023jot,Franciolini:2023pbf,Franciolini:2023wjm,Inomata:2023zup,Cai:2023dls,Wang:2023ost,Liu:2023ymk,Unal:2023srk,Figueroa:2023zhu,Yi:2023mbm,Zhu:2023faa,Firouzjahi:2023lzg,Li:2023qua,You:2023rmn,Balaji:2023ehk,HosseiniMansoori:2023mqh,Zhao:2023joc,Liu:2023pau,Yi:2023tdk,Bhaumik:2023wmw,Choudhury:2023hfm,Yi:2023npi,Harigaya:2023pmw,Basilakos:2023xof,Jin:2023wri,Cannizzaro:2023mgc,Zhang:2023nrs,Liu:2023hpw,Choudhury:2023fwk,Tagliazucchi:2023dai,Basilakos:2023jvp,Inomata:2023drn,Li:2023xtl,Domenech:2023dxx,Gangopadhyay:2023qjr,Cyr:2023pgw,Lozanov:2023rcd,Madge:2023dxc} for subsequent works). If that is the case, these results hint  at new physics during cosmic inflation and possibly a significant abundance of Primordial Black Holes (PBHs).

There is strong evidence of cosmic inflation from observations of the Cosmic Microwave Background (CMB) \cite{Planck:2018jri}, with a measured amplitude of the power spectrum of primordial curvature fluctuations of about $10^{-9}$ at Mpc -- Gpc scales. Yet, to reproduce the reported amplitude of the GW background, the spectrum of primordial fluctuations should have a peak amplitude of about $10^{-2}$ at roughly parsec scales \cite{NANOGrav:2023hvm}. Nevertheless, there are plenty of inflationary models capable of enhancing the primordial spectrum of curvature fluctuations on small scales, such as ultra-slow-roll (USR) \cite{Kinney:2005vj, Martin:2012pe, Cheng:2018qof, Byrnes:2018txb, Pi:2022zxs, Cheng:2023ikq, Kawai:2021edk}, sound speed resonance (SSR) \cite{Cai:2018tuh, Chen:2019zza}, multi-field inflation \cite{Kawasaki:1997ju, Kawasaki:2012wr, Pi:2017gih, Kawai:2022emp}, and the curvaton scenario \cite{Pi:2021dft,Ferrante:2022mui,Meng:2022low,Chen:2023lou} among others (see \textit{e.g.} Ref.~\cite{Ozsoy:2023ryl} for a review and references therein). 
Most interestingly, though, such a large variance for primordial fluctuations implies that some fluctuations may be large enough to end up forming PBHs  \cite{Zeldovich:1967lct,Hawking:1971ei,Hawking:1974rv,Carr:1974nx,Meszaros:1974tb,Hawking:1975vcx,Carr:1975qj,Khlopov:1985jw} (for recent reviews on PBHs see also Refs.~\cite{Khlopov:2008qy,Sasaki:2018dmp,Carr:2020gox,Green:2020jor,Escriva:2022duf}). 

PBHs associated with the induced GW interpretation of the PTA data (assuming a radiation dominated universe) apparently have a mass around $10^{-4}M_\odot-10^{-2}M_\odot$ \cite{Franciolini:2023pbf,Inomata:2023zup}, where $M_\odot$ is a solar mass. However, despite important uncertainties in the calculations of the PBH abundance, one of the issues pointed out in Ref.~\cite{NANOGrav:2023hvm} is that the amplitude of the primordial spectrum given by the $2\sigma$ contours from PTA analysis tends to predict too many PBHs in that mass range (assuming Gaussian primordial fluctuations and a radiation dominated very early universe). However, while the induced GW spectrum depends quadratically on the power spectrum, PBH abundance is exponentially sensitive to it. This means, on the one hand, that finding induced GWs might not necessarily imply a significant fraction of PBHs. For example, Ref.~\cite{Inomata:2023zup} argues that a small fraction of the available parameter space could explain the planet-mass microlensing events seen by OGLE \cite{Mroz:2017mvf,Niikura:2019kqi}. This is an interesting possibility as it may also give a detectable GW background from the mergers of PBH binaries \cite{Inomata:2023zup}. On the other hand, it also means that relatively small changes in the statistical nature of primordial fluctuations could significantly suppress the PBH abundance, as is the case of negative non-Guassianity of primordial fluctuations\footnote{Induced GWs are not very sensitive to the amount of non-Gaussianity, see \textit{e.g.} Refs.~\cite{Cai:2018dig,Unal:2018yaa,Atal:2021jyo,Adshead:2021hnm,Abe:2022xur,Yuan:2023ofl,Li:2023xtl}.} \cite{Franciolini:2023pbf, Wang:2023ost, Figueroa:2023zhu, Choudhury:2023fwk,Liu:2023ymk,Li:2023qua} (which might not be straightforward to obtain in single field models of inflation \cite{Pi:2022ysn,Firouzjahi:2023xke}) and a period of the universe deviating from standard radiation domination \cite{Domenech:2020ers,Balaji:2023ehk,Zhao:2023joc,Liu:2023pau,Harigaya:2023pmw, Choudhury:2023fjs}.

 In this paper, we take an agnostic approach. Since the induced GWs and the PBHs corresponding to PTA frequencies were generated and formed when the temperatures in the universe were above $100\,{\rm MeV}$ (a period of the universe which has not been probed directly), we do not assume any particular value of the equation of state (EoS) of the universe, commonly denoted $w$. For comparison, note that a successful Big Bang Nucleosynthesis (BBN) requires the universe to be radiation dominated ($w=1/3$) and thermalized for $T\gtrsim4 \,{\rm MeV}$ \cite{Kawasaki:1999na,Kawasaki:2000en,Hannestad:2004px,Hasegawa:2019jsa} (or at redshifts greater than $10^{9}$). Other values of the EoS are tied to different models of reheating (see, \textit{e.g.}, Ref.~\cite{Allahverdi:2020bys} for a review). For example, the coherent oscillations of a scalar field yield $w\sim 0$, and in quintessential inflation scenarios, one has a period with $w=1$ after inflation \cite{Spokoiny:1993kt,Peebles:1998qn,Brax:2005uf,Hossain:2014xha}. In this way, since the induced GW spectrum depends on $w$, it serves as a probe of the content of the primordial universe \cite{Assadullahi:2009nf,Inomata:2019zqy,Inomata:2019ivs,Inomata:2020lmk,Papanikolaou:2020qtd,Domenech:2020ssp,Domenech:2021wkk,Dalianis:2020gup,Hajkarim:2019nbx,Bhattacharya:2019bvk,Domenech:2019quo,Domenech:2020kqm,Dalianis:2020cla,Abe:2020sqb,Witkowski:2021raz,Balaji:2022dbi,Lozanov:2023aez,Lozanov:2023knf}. 
 
 We would like to emphasize that while this approach has also been considered for PTAs in Refs.~\cite{Domenech:2020ers,Zhao:2023joc,Liu:2023pau,Harigaya:2023pmw,Choudhury:2023fjs}, we improve previous analysis by performing the most thorough study on both the induced GWs and PBHs related to PTAs.  Requiring no PBH overproduction, we obtain that $0.42\lesssim w\lesssim 0.87$. And, using CMB bounds on the effective number of relativistic species (which sets upper bounds on the peak GW frequency), we conclude that the allowed range of the typical PBH masses is given by $2\times10^{-5} M_\odot\lesssim M_{\rm PBH}\lesssim 0.09 M_\odot$. Interestingly, the mass range allowed by OGLE for PBHs, namely $[1.6\times 10^{-6}M_\odot, 6.9\times10^{-5}M_\odot ]$  \cite{Niikura:2019kqi}, overlaps slightly with the range we obtained for $0.42\lesssim w\lesssim 0.50$. There are also candidate merger events in the LVK data of compact objects with a mass of about $O(0.1)M_\odot$  \cite{Clesse:2020ghq,Morras:2023jvb,LIGOScientific:2022hai,Prunier:2023cyv}. 
 For simplicity, we do not include non-Gaussianity of primordial fluctuations. But, as future work, it would be interesting to study both effects simultaneously.

This paper is organized as follows. In \S~\ref{sec: SIGW}, we review the induced GW spectrum generated during the general EoS stage for a log-normal primordial spectrum. In \S~\ref{sec: Bayes}, we present the results of our Bayesian analysis for a log-normal and a Dirac delta primordial spectra. We conclude our work in \S~\ref{sec: c&d} with further discussions. Lastly, we provide details on the PBH calculations in App.~\ref{sec: PBH} and on the CMB bounds on extra relativistic species in App.~\ref{app:CMB}.

\section{Induced Gravitational Waves in a general \texorpdfstring{$w$}{} stage}
\label{sec: SIGW}

There is an inevitably secondary production of GW  induced by the curvature perturbation \cite{Tomita,Matarrese:1992rp,Matarrese:1993zf,Carbone:2004iv}. The spectrum of such induced GWs \cite{Ananda:2006af,Baumann:2007zm,Saito:2008jc,Saito:2009jt,Bugaev:2009zh,Bugaev:2010bb} (see Refs.~\cite{Domenech:2021ztg,Domenech:2023jve} for recent reviews) depends on the spectrum of the curvature perturbation as well as on the EoS of the universe at the time of generation \cite{Domenech:2019quo, Domenech:2021ztg, Witkowski:2022mtg}. 
We assume that around a specific wavenumber $k_\mathrm{p}$, there is a peak in the primordial power spectrum of curvature fluctuations ${\cal R}$. Inspired by the PBH mass function \cite{Dolgov:1992pu,Green:2016xgy,Carr:2017jsz,Gorton:2024cdm}, model-independent reconstruction \cite{Kimura:2021sqz,Wang:2022nml,Sharma:2024whg}, and the inflationary model building \cite{Kawasaki:1997ju,Kinney:2005vj,Martin:2012pe,Kawasaki:2012wr,Pi:2017gih,Cheng:2018qof,Byrnes:2018txb,Cai:2018tuh,Chen:2019zza,Kawai:2021edk,Pi:2022zxs,Cheng:2023ikq}, we parametrize such a peaked spectrum with a log-normal function given by \cite{Pi:2020otn}
\begin{equation}
    \label{eq:log-normal perturbation}
    \mathcal{P}_{\mathcal{R}}(k)=\frac{{\cal A}_{\mathcal{R}}}{\sqrt{2 \pi} \Delta} \exp \left[-\frac{\ln ^2\left(k / k_{\mathrm{p}}\right)}{2 \Delta^2}\right],
\end{equation}
where $\mathcal{A}_{\mathcal{R}}$ is the amplitude, $\Delta$ is the logarithmic width, and $k_{\rm p}$ is the peak scale. The commonly used $\delta$-function power spectrum corresponds to $\Delta\to0$ limit of \eqref{eq:log-normal perturbation}.

Since  induced GWs (and  PBHs) were produced during an era dominated by a perfect fluid with a constant $w$, there is another important scale besides the peak scale $k_{\rm p}$, namely the scale of reheating $k_{\rm rh}$. For simplicity, we take an almost instantaneous transition from a general $w$ to the standard radiation domination with $w=1/3$.\footnote{Note that an exact instantaneous transition leaves some high frequency oscillatory patterns in the induced GW spectrum, as shown in Ref.~\cite{Altavista:2023zhw}. However, we believe these are an artifact of the instantaneous transition and would quickly disappear as one considers a finite duration. Also, when the instantaneous transition goes from $c_s^2=w=0$ to $w=1/3$, there is an enhanced production of the induced GWs \cite{Inomata:2019ivs,Inomata:2019zqy}. We do not consider this case here.
} In this case, the reheating scale corresponds to the scale that entered the horizon at the moment of reheating, that is $k_{\rm rh}=a_{\rm rh}H_{\rm rh}$ where $a_{\rm rh}$ and $H_{\rm rh}$ respectively are the scale factor and the Hubble parameter at reheating. For further simplicity, we also assume that the frequency corresponding to the reheating scale evaluated today, namely $f_{\rh}={k_{\rh}}/({2\pi a_{0}})$,
is smaller than the lowest frequency bin of PTA data. In other words, we impose $f_\rh<f_{\rh,\mathrm{max}}\approx 2\times10^{-9}\,{\rm Hz}$. By doing so, we do not need the explicit expression of the induced GW spectrum for $f<f_\rh$ as it never enters the observational range. Also, it is important to note that for $w=1/3$, the “reheating” scale is meaningless, as there is no reheating. In that case, $f_\rh$ can be considered an arbitrary pivot scale.

If we let $f_{\rm rh}$ vary, we would find two possibilities: either $f_{\rm rh}$ is well inside or below the PTA frequencies. The former is not so interesting as one mostly recovers the results of standard radiation domination $w=1/3$, already studied in \cite{NANOGrav:2023hvm}. The latter is bounded from below by BBN constraints on the reheating temperature, $T\gtrsim4 \,{\rm MeV}$ \cite{Kawasaki:1999na,Kawasaki:2000en,Hannestad:2004px,Hasegawa:2019jsa}. From now on, we consider a fixed $f_{\rm rh}$ because our main focus is only to understand the effects of different values of $w$. For practical purposes, we set
\begin{align}\label{eq:frhstar}
  f_{\rh,\star}\approx 4\times10^{-10}~\mathrm{Hz}\,,
\end{align}
which corresponds to $T_{\rh,\star} =20\,{\rm MeV}$, assuming the standard model of particle physics. Such an arbitrary choice is roughly in the middle of the lowest frequency of PTAs ($\sim 10^{-9}\,{\rm Hz}$) and the frequency corresponding to the lowest reheating temperature ($\sim 10^{-10}\,{\rm Hz}$). Our main results do not depend significantly on the choice of the reheating scale, yielding at most some ${\cal O}(1)$ factor variations, as we show later. 
For future reference, the temperature after reheating $T_{\rh}$ is given by
\begin{equation}\label{eq:reheating temperature}
    T_{\rh}= 110\,\text{MeV} ~\left(\frac{f_\rh}{f_{\rh,\text{max}}}\right)~\left(\frac{g_{\rho,\rh}}{10.75}\right)^{-1 / 2}\left(\frac{g_{s,\rh}}{10.75}\right)^{1 / 3}\,,
\end{equation}
where $g_{\rho,\rh}$ and $g_{s,\rh}$ respectively are the effective degrees of freedom in the energy density $\rho$ and entropy density $s$ of the radiation fluid evaluated at the reheating temperature, $T_{\rm rh}$. For the values of $g_{\rho,\rh}$ and $g_{s,\rh}$ in the standard model of particles see, \textit{e.g.}, Ref.~\cite{Saikawa:2018rcs}.  Note that in Eq.~\eqref{eq:reheating temperature} there is an implicit dependence on $T_\rh$ inside $g_{\rho,\rh}$ and $g_{s,\rh}$, which we neglected as we are considering the small range $4\,{\rm MeV}<T_\rh<110\,{\rm MeV}$.

The spectral energy density of the induced GWs produced during an era dominated by a perfect fluid with a constant $w>0$ is given by \cite{Domenech:2019quo, Domenech:2021ztg,Witkowski:2021raz, Witkowski:2022mtg}
\begin{equation}
    \label{eq: SIGW spectrum}
    \Omega_{\mathrm{GW},
    \rm rh}(k>k_\rh)=\left(\frac{k}{k_{\mathrm{rh}}}\right)^{-2 b} \int_0^1 \mathrm{~d} d \int_1^{\infty} \mathrm{d} s ~\mathcal{T}_w(d, s) \,\mathcal{P}_\Rcal\left(\frac{k}{2}(s+d)\right) \mathcal{P}_\Rcal\left(\frac{k}{2}(s-d)\right),
\end{equation}
where
\begin{align}
b=\frac{1-3w}{1+3w}\,,
\end{align}
and $\mathcal{T}_w(d, s)$ is the general kernel for modes reentering horizon before reheating,  given, \textit{e.g.}, in Refs.~\cite{Witkowski:2021raz,Domenech:2021ztg}. In Eq.~\eqref{eq: SIGW spectrum}, $d$ and $s$ are related to the internal momenta of the scalar modes. For the first derivation of semi-analytical formulas in radiation domination, see Refs.~\cite{Espinosa_2018,Kohri:2018awv}. Note that in \eqref{eq: SIGW spectrum}, the factor $\left({k}/{k_{\mathrm{rh}}}\right)^{-2 b}$ comes from the dilution of the energy density of subhorizon GW modes, which decay as radiation (\textit{i.e.} $\rho_{\GW} \propto a^{-4}$), as compared to background energy density (that is $\rho_{w} \propto a^{-3(1+w)}$). Using entropy conservation, the current spectral density of induced GWs reads \cite{Inomata:2018epa}
\begin{align}\label{eq:omegatoday}
    \Omega_{\GW,0}(k)h^2&=\frac{g_{\rho, \rh}}{g_{\rho, 0}}\left(\frac{g_{s, 0}}{g_{ s, \rh}}\right)^{4 / 3}\Omega_{\GW,\rm rh}(k)\Omega_{r,0}h^2\notag\\
    &=3.40 \times 10^{-5}\left(\frac{g_{ s,\rh}}{10.75}\right)^{-4 / 3}\left(\frac{g_{\rho,\rh}}{10.75}\right)\left(\frac{\Omega_{r, 0} h^2}{2.47 \times 10^{-5}}\right) \Omega_{\mathrm{GW},\rm rh}(k)\,,
\end{align}
where $\Omega_{r, 0} h^2$ is the reduced density ratio of radiation today given by Planck \cite{Planck:2018vyg}. From Eq.~\eqref{eq: SIGW spectrum} we see that the amplitude of the induced GW spectrum depends on the amplitude ${\cal A}_{\cal R}$ of the power spectrum \eqref{eq:log-normal perturbation}, the equation of state $w$ and the ratio of scales $k_{\rm p}/k_{\rm rh}$.

Before we get into the details and results of the Bayesian analysis, it is useful to have an analytic understanding of the spectrum of induced GWs in the low frequency tail, also called infra-red (IR) tail, which corresponds to $k_{\rm rh}\ll k\ll k_{\rm p}$. This is because, in the end, the power-law behavior of the IR tail mostly fits the PTA data. We now discuss the case of a Dirac delta and a log-normal primordial spectrum separately.

\subsection{Dirac delta peak \label{subsec:diracdelta}}

We start with a Dirac delta primordial spectrum given by
\begin{align}\label{eq:diracdelta}
\Pcal_\Rcal(k)=\Acal_\Rcal \delta(\ln({k/k_{\rm p}}))\,.
\end{align}
In this case, the spectrum of induced GW after inserting Eq.~\eqref{eq:diracdelta} into Eq.\eqref{eq: SIGW spectrum} exactly reads
\begin{equation}\label{eq: Dirac delta}
    \Omega_{\mathrm{GW}, w}^{(\delta)}(k)=2~\Acal^2_\Rcal\left(\frac{k_{\rm p}}{k_{\mathrm{rh}}}\right)^{-2 b} \left[\left(\frac{k}{k_{\rm p}}\right)^{-2-2 b} \mathcal{T}_w(s=2k/k_{\rm p},d=0)\Theta(2-k/k_{\rm p})\right].
\end{equation}
We are interested in the IR tail of the induced GWs, which corresponds to the limit $k\ll k_{\rm p}$ of Eq.~\eqref{eq: Dirac delta}. After a Taylor expansion, one finds \cite{Domenech:2020kqm,Domenech:2021ztg}
\begin{align}\label{eq:IRDIRACDELTA}
 \Omega_{\mathrm{GW}, {w}}^{(\delta)}(k\ll k_{\rm p})\approx F[b,c_s]\,\frac{{\cal A}_{\cal R}^2}{c_s^4}\left(\frac{k_{\rm p}}{k_{\rm rh}}\right)^{-2b}\left(\frac{k}{k_{\rm p}}\right)^{2-2|b|}&\,,
\end{align}
where $c_s^2=w$ is the sound speed of fluctuations, and $F[b,c_s]$ is a function of $b$ and $c_s$, whose concrete form is irrelevant for the following discussion.\footnote{ The function  $F[b,c_s]$ in Eq.~\eqref{eq:IRDIRACDELTA} is given by
\begin{align}\label{eq:fbcs}
F[b,c_s]\equiv
\frac{4^b(2+b)^2}{3(1+b)^{2(1+b)} }
    \Gamma^2[\tfrac{3}{2}+b]\left\{
    \begin{aligned}
        &\frac{\Gamma^2[\tfrac{3}{2}+b]}{c_s^{4 b}\sin^2(b\pi)\Gamma^2[2+b]}& (b<0\,\,;\,\,w>1/3)\\&
        \frac{\left(1+b+b^2\right)^2}{4 \pi 
        b^2 (1+b)^4}& (b>0\,\,;\,\,w<1/3)
    \end{aligned}
 \right. \,.
\end{align}}

Using Eq.~\eqref{eq:IRDIRACDELTA} we may study degeneracies among model parameters, which are relevant for the subsequent data analysis.  First, we note that the function $F[b,c_s]$ yields at most ${\cal O}(1-10)$ variations for $1>b>-1/2$ (or $0<w<1$) and $c_s^2=w$. For instance, for $b=1$ and $b=-1/2$ one respectively finds $F[b,c_s]\approx 0.06$ and $F[b,c_s]\approx 0.95$. This shows that the IR tail has a mild dependence on $w$ through the function $F[b,c_s]$. But, since the factor $c_s^{-4}$ in Eq.~\eqref{eq:IRDIRACDELTA} yields a much larger change for varying $w$,  we neglect the dependence on $F[b,c_s]$ in the following discussion. Thus, the amplitude of the IR tail is then mostly dependent on the combination (for a fixed $k$) given by
\begin{align}\label{eq:IRdegeneracy}
\frac{{\cal A}_{\cal R}}{c_s^2}\left(\frac{k_{\rm p}}{k_{\rm rh}}\right)^{-b}\left(\frac{k}{k_{\rm p}}\right)^{1-|b|}\approx {\rm constant}.
\end{align}
For $b=0$ ($w=1/3$), one recovers the results of Ref.~\cite{Balaji:2023ehk}.

Since the reported amplitude of the GW background is fairly large, the peak of the induced GW spectrum cannot lie too far from the PTA window; otherwise, it might conflict with CMB constraints on the effective number of relativistic species. For this reason, it is also instructive to study the amplitude of the induced GW spectrum at the higher end of the IR tail, or the “bump” of the GW spectrum. Based on the results for $w=1/3$ of Ref.~\cite{Balaji:2023ehk}, we find that the bump is located to a good approximation at
\begin{align}
k_{\rm bump}\approx \frac{1}{\sqrt{2}}c_sk_{\rm p}\,,
\end{align}
with an amplitude given by
\begin{align}\label{eq:GWbump}
\Omega_{{\rm GW},w}^{{(\delta)},\rm bump}\approx \frac{C[b]}{c_s^{2(1+b)}} \left(1 - \frac{c_s^2}{8}\right)^2{\cal A}_{\cal R}^2\left(\frac{k_{\rm p}}{k_{\rm rh}}\right)^{-2b}
\end{align}
where we introduced for compactness the function $C[b]$.\footnote{The function $C[b]$ in Eq.~\eqref{eq:GWbump} is given by
\begin{align}
C[b]=&\frac{ 7^{2 b} (1+b)^{-2 (3+b)} (2+b)^2 (18+25 b)^2 \,\Gamma^4
   [b+\tfrac{3}{2}] }{3\times 2^{11+5 b}\, \Gamma^2[1+b]}\nonumber\\&\times\left\{1+\left(\cot (b\pi )+\frac{ 2^b\sqrt{\pi }  \csc (\pi  b)}{\Gamma[1-b] \Gamma[b+\tfrac{1}{2}]}\left(\frac{3\times
   8^{1+b} b (2+b)}{7^{b}(1+2 b) (18+25 b)}-\,
   _2F_1\left[-b,1+b;1-b;\tfrac{1}{8}\right]\right)\right)^2\right\}\,.
\end{align}}
Such a function is bounded by $0.5>C[b]>0.1$ for $1>b>-1/2$ ($0<w<1$). Thus, the induced GW spectrum has a similar amplitude and position of the bump (for a fixed $k$) when 
\begin{align}\label{eq:bumpdegeneracy}
\frac{{\cal A}_{\cal R}}{c_s^{1+b}}\left(\frac{k_{\rm p}}{k_{\rm rh}}\right)^{-b}\approx {\rm constant} \qquad {\rm and} \qquad \frac{1}{\sqrt{2}}c_sk_{\rm p}\approx {\rm constant}\,.
\end{align}
From Eqs.~\eqref{eq:IRdegeneracy} and \eqref{eq:bumpdegeneracy}, we see that increasing the value of $w$ allows for a decrease in the value of ${\cal A}_{\cal R}$. For $b<0$ ($b>0$), we see that increasing the value of $k_{\rm p}$ (or lowering the value of $k_{\rm rh}$) leads to a decrease (increase) to the necessary value of ${\cal A}_{\cal R}$. We will encounter such behavior in the Bayesian analysis of \S~\ref{sec: Bayes}.

\subsection{Log-normal peak \label{subsec:lognormal}}

The case of the log-normal peak \eqref{eq:log-normal perturbation} has not been studied analytically for general $w$. Analytical approximations for the case of radiation domination have been derived in Ref.~\cite{Pi:2020otn}. Nevertheless, we may get some intuition if we focus on the far IR tail of the induced GW spectrum. In this case, most of the results of the Dirac delta carry over to the log-normal case with few changes. In particular, in the far IR, there is a change of slope from the $k^{2-2|b|}$ of the Dirac delta to $k^{3-2|b|}$ for the log-normal \cite{Cai:2019cdl,Domenech:2020kqm}. We follow Ref.~\cite{Pi:2020otn} to provide more precise estimates to derive some approximations in the far IR limit. Concretely, we perform the integral Eq.~\eqref{eq: SIGW spectrum} in the limit $k\ll k_{\rm p}\Delta e^{-\Delta^2}$, which corresponds to the regime where $s\gg1$ and $d\ll1$ and where the log-normal power spectrum may be considered as a very sharp peak. By doing so, we obtain that
\begin{align}\label{eq:IRfinite}
 \Omega_{\mathrm{GW}, w}^{(\Delta)}(k\ll k_{\rm p}\Delta e^{-\Delta^2})\approx \Omega_{\mathrm{GW},w}^{(\delta)}(k\ll k_{\rm p})\times\left(\frac{k}{k_{\rm p}}\right)\frac{1}{2\sqrt{\pi}\Delta}\left\{
 \begin{aligned}
&e^{\left(\tfrac{3}{2}+2b\right)^2\Delta^2}& (b<0\,\,;\,\,w>1/3)\\&
e^{\tfrac{9}{4}\Delta^2}& (b\geq0\,\,;\,\,w\leq1/3)
 \end{aligned}
 \right.\,,
\end{align}
where $\Omega_{\mathrm{GW}, w}^{(\delta)}(k\ll k_{\rm p})$ is given in Eq.~\eqref{eq:IRDIRACDELTA}. It is worth mentioning that, while performing numerical integrations, we noticed that the approximations of Ref.~\cite{Pi:2020otn} for $w=1/3$ present some variations in the case of general $w$. In particular, the precise position of the change of slope from $k^{2-2|b|}$ to $k^{3-2|b|}$ depends on the value of $w$.\footnote{For example, for a narrow log-normal peak ($\Delta<0.1$) there is a good approximation to the log-normal case for $k<\Delta\, k_{\rm p}$ and $w=1/3$ which is given by \cite{Pi:2020otn,Domenech:2021ztg}
\begin{equation}
    \label{eq: Finite width correction}
    \Omega_{\mathrm{GW}}^{(\Delta)}=\operatorname{Erf}\left[\frac{1}{\Delta} \sinh ^{-1} \frac{k}{2 k_{\mathrm{p}}}\right] \Omega_{\mathrm{GW}}^{(\delta)}. 
\end{equation}
This equation should be derived more precisely for general values of $w$ by including the dependence on $w$ in the function in front of $ \Omega_{\mathrm{GW}}^{(\delta)} $.} While interesting, this is out of the scope of this paper. We will present detailed analytical approximations in subsequent works.

\section{Bayesian analysis and results}
\label{sec: Bayes}

We perform the Bayesian analysis with the package \href{https://andrea-mitridate.github.io/PTArcade/}{\textsc{PTArcade}} \cite{Mitridate:2023oar} using the ceffyl model \cite{lamb2023need}. Also, as in the NANOGrav analysis \cite{NANOGrav:2023hvm}, we only consider the first 14 frequency bins as signals for the GW background \cite{NANOGrav:2023hvm}. At the same time, the pulsar-intrinsic red noise is obtained considering all frequency bins. To derive the likelihood of the amplitude of the GW background signal, say $P(\Theta \mid \mathcal{D}, \mathcal{M})$, we need to obtain the induced GW spectra in various target models $\mathcal{M}$ with different model parameters $\Theta$ for analyzing the data $\cal D$. We do so as follows.

\begin{figure}
\includegraphics[width=0.6\linewidth]{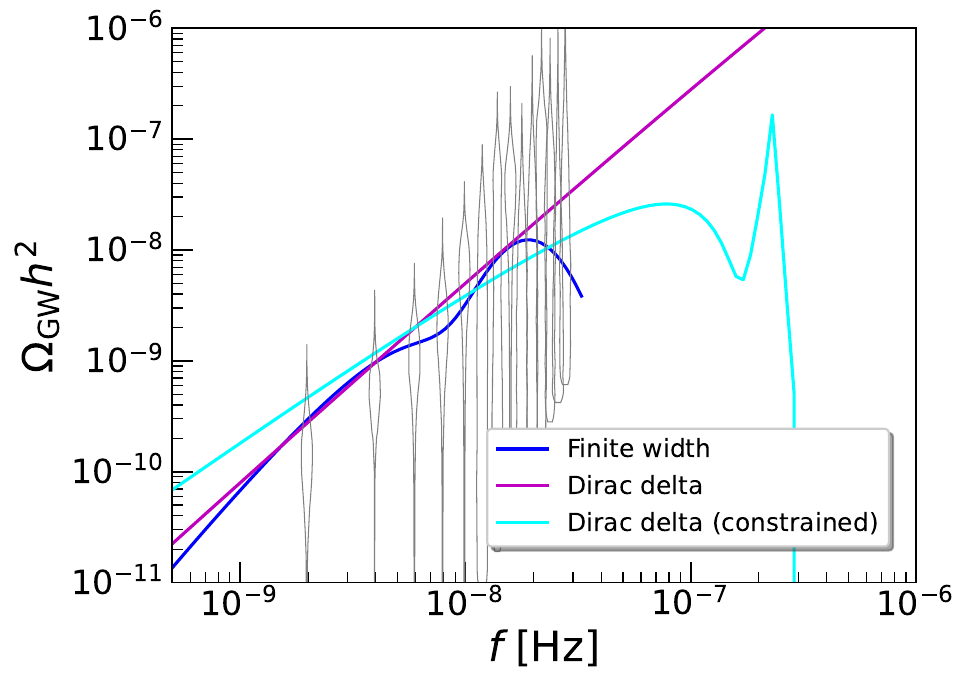}
    \caption{Best fit induced GW spectra for different models using the first 14 frequency bins of NANOGrav data (the gray violins). The best fit parameters for the Dirac delta \eqref{eq:diracdelta} read $[w, \log_{10} {\cal A}, \log_{10} f_{\rm p}]=[0.35, 0.98, -3.79]$, shown in purple. Including the bound from no PBH overproduction the best fit moves to $[w, \log_{10} {\cal A}, \log_{10} f_{\rm p}]=[0.59, -2.02, -6.81]$ which we show with a cyan line. For the log-normal \eqref{eq:log-normal perturbation} we find $[w, \log_{10} {\cal A}, \log_{10} f_{\rm p}, \Delta]=[0.31, -1.60, -7.79, 0.49]$.}
    \label{fig:fit_result}
\end{figure}

We use the analytical formula, Eq.~\eqref{eq: Dirac delta}, for the Dirac delta case. For the log-normal, we use the \href{https://github.com/Lukas-T-W/SIGWfast/releases}{\textsc{SIGWfast}} \cite{Witkowski:2022mtg} package for an efficient numerical calculation of the energy spectra of induced GWs for general $w$. Since the role of sound speed has already been discussed in Ref.~\cite{Balaji:2023ehk} (although for fixed $w=1/3$), we only consider the adiabatic perfect fluid case with $c_s^2=w$ in our sampling. We checked that the case of $c_s^2=1$ yields similar results for the GW spectrum. In our analysis, we also consider \textit{a posteriori} the PBH counterpart and CMB constraints on the effective number of relativistic species. For details on the calculations of the PBH abundance and CMB constraints, we refer the reader respectively to Apps.~\ref{sec: PBH} and \ref{app:CMB}. We show the GW spectrum for the best fit parameters of different models in Fig. \ref{fig:fit_result}.
 The results of the Bayesian analysis are shown in Figs.~\ref{fig:bayesiandelta} and \ref{fig:bayesianbroad}, and our priors and  95\% posterior confidence intervals are presented in Tabs.~\ref{tab:Dirac Delta Uncertainty} and \ref{tab:Finite Delta Uncertainty}.

For clarity, we summarize below the parametrization used in our analysis. We compute the induced GW spectrum \eqref{eq: SIGW spectrum} as
\begin{equation}\label{eq:omegarh}
    \Omega_{\GW,\rm rh}(f)= {\cal A}^2 \left(\frac{f}{f_{\rh,\star}}\right)^{-2b}\int_0^1 \mathrm{~d} d \int_1^{\infty} \mathrm{d} s\, \mathcal{T}_w(d, s)\, G_\Delta\left(\frac{f(s+d)}{2 f_{\rm p}}\right) G_\Delta\left(\frac{f(s-d)}{2 f_{\rm p}}\right)\,,
\end{equation}
where $f_{\rh,\star}\approx 4\times10^{-10}~\mathrm{Hz}$ (as explained around Eq.~\eqref{eq:frhstar}),
\begin{align}
G_\Delta(x)=\frac{1}{\sqrt{2\pi}\Delta}\exp \left[-\frac{\ln^2x}{2 \Delta^2}\right]\,.
\end{align}
and we re-defined
\begin{align}\label{eq:calA}
{\cal A}\equiv \Acal_\Rcal \left(\frac{f_\rh}{f_{\rh,\star}}\right)^{b}\,.
\end{align}
With the above redefinition, Eq.~\eqref{eq:calA}, the constraints we find in the variable ${\cal A}$ can be easily translated to constraints in $\Acal_\Rcal$ for a given value of $f_\rh$. For instance, for $f_{\rh,\rm min}\sim 10^{-10}$ (roughly corresponding to $T_{\rh}\sim 4\,{\rm MeV}$) and $f_{\rh,\rm min}\sim 10^{-9}$ (roughly corresponding to $T_{\rh}\sim 100\,{\rm MeV}$) we respectively have that ${f_{\rh,\rm  min}}/{f_{\rh,\star}}\sim 4/20\sim 1/5 $ and ${f_{\rh,\rm  max}}/{f_{\rh,\star}}\sim 100/20\sim 5$. In fact, since for the Dirac delta case we eventually find that no PBH overproduction requires $w>0.42$, the variation in reheating frequency can only amount at most to a factor $\sim 2$ for $f_{\rh,\rm min}$ (or $\sim 1/2$ for $f_{\rh,\rm max}$) taking the maximum value of $|b|$ in that range, namely $b=-1/2$. Thus, our results for a fixed $f_{\rh}=f_{\rh,\star}$ already provide good intuition for the general case of $f_{\rh,\rm min}<f_{\rm rh}<f_{\rh,\rm max}$. We proceed to discuss our results for the Dirac delta and log-normal cases.

\subsection{Results for Dirac delta peak  \label{subsec:resultsDirac}}

\begin{figure}
\includegraphics[width=0.6\linewidth]{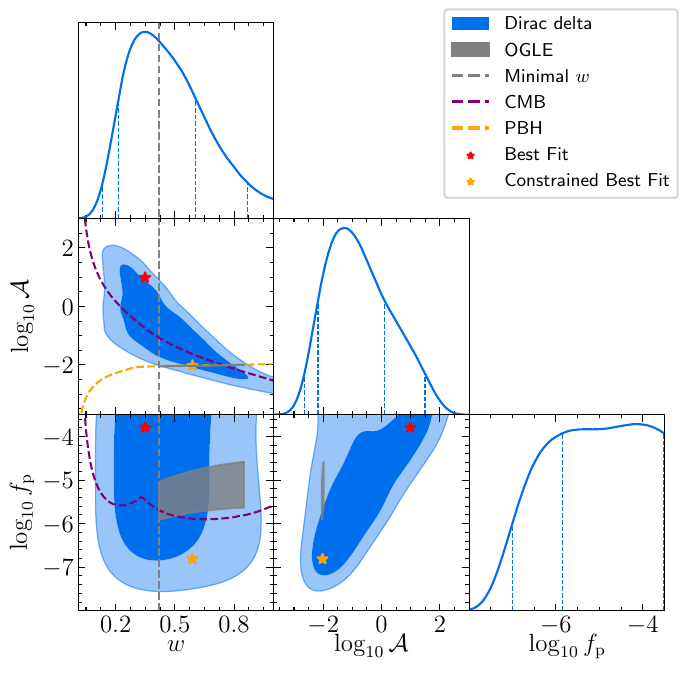}
    \caption{Posterior distribution inferred from the NANOGrav spectrum probability distribution function for the Dirac delta primordial spectrum \eqref{eq:diracdelta}. In orange and purple dashed lines, we respectively show the constraints from no PBH overproduction (see App.~\ref{sec: PBH}) and the effective number of relativistic species in the CMB (see App.~\ref{app:CMB}). The regions above these lines are excluded from the parameter space. The gray shaded region shows the corresponding parameters space for the 95\% CL mass range in OGLE 5-year \cite{Niikura:2019kqi}, including the PBH abundance. The stars show the best fit parameters before (red) and after (orange) considering PBH and CMB bound. We see that there is still a small parameter space to explain OGLE microlensing events and PTA signal simultaneously. The blue dashed lines in the marginalized distributions show the 68\% and 95\% HDPI in the marginalized distributions.
    }
    \label{fig:bayesiandelta}
\end{figure}

\begin{figure}
\includegraphics[width=0.6\linewidth]{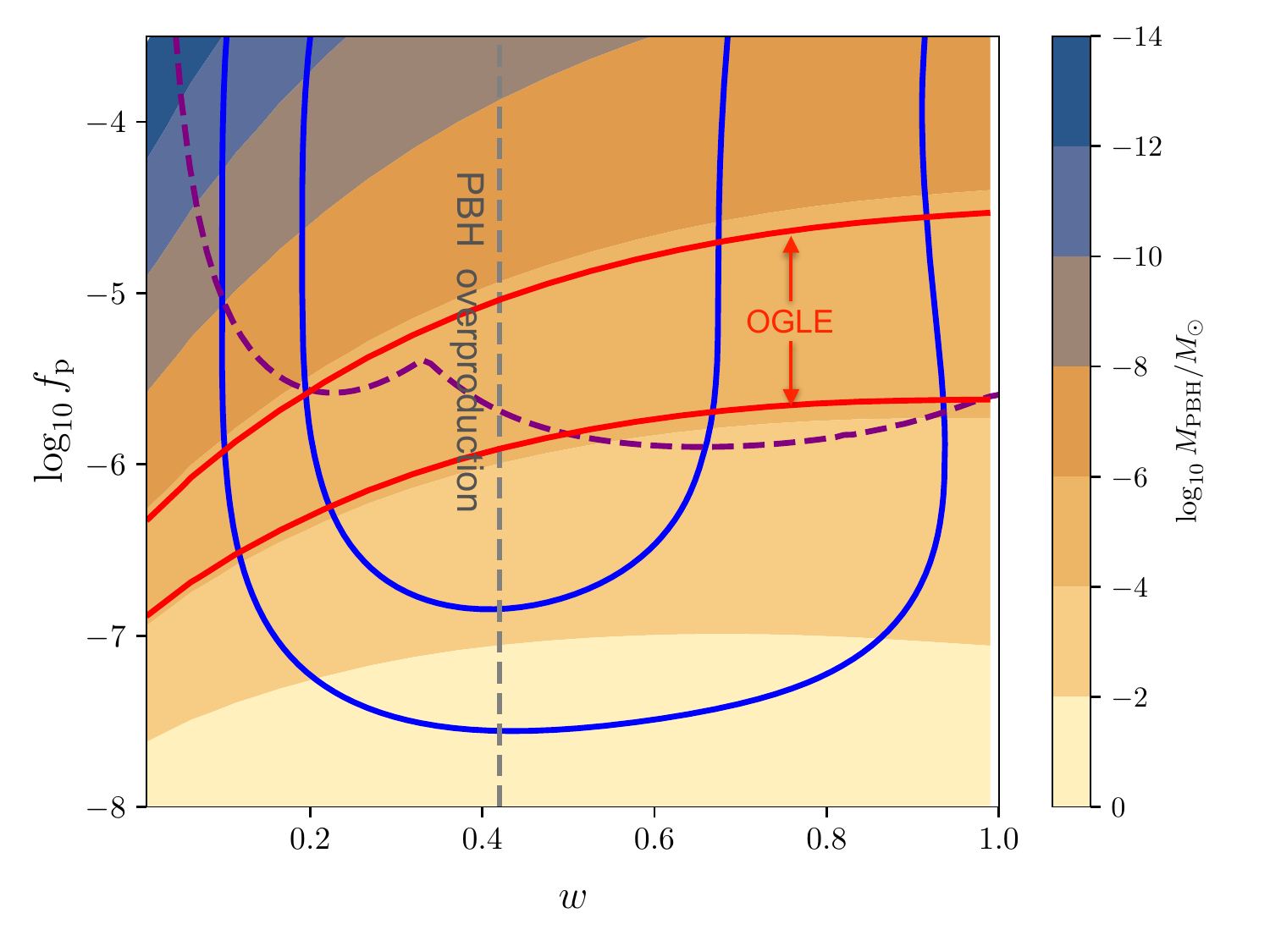}
    \caption{Peak of the mass spectrum, $\log_{10} M/M_\odot$, of the PBH counterpart to the induced GWs with respect to different EoS $w$ and peak frequency $\log_{10} f_{\rm p}$. Red lines show the minimum and the maximum PBH mass (respectively $M_{\rm min}=1.6\times10^{-6} M_\odot$ and $M_{\rm max}=6.9\times10^{-5} M_\odot$) compatible with the microlensing events seen by OGLE \cite{Niikura:2019kqi}. Gray lines indicate the EoS $w$ in which there are too many PBHs (left of the dashed line) and too few PBHs to be observed (right of the dot-dashed line). The blue lines show the 1-$\sigma$ and 2-$\sigma$ of NANOGrav data, and the purple line shows the CMB bound as Fig. \ref{fig:bayesiandelta}.
    }
    \label{fig:peak mass}
\end{figure}

We present the priors used and the 95\% highest probability distribution interval (HPDI) of the marginalized distributions in the Dirac delta case in Tab.~\ref{tab:Dirac Delta Uncertainty}. In Fig.~\ref{fig:bayesiandelta}, we show the posterior distributions with the NANOGrav dataset. In the same figure, we present with an orange dashed line the upper bound on the amplitude of primordial fluctuations from the requirement that PBHs constitute $100\%$ of the dark matter (\textit{i.e.}, above the orange line PBHs are overproduced and inconsistent with observations). The gray dashed vertical line then shows the lower bound on the equation of state $w$ with no PBH overproduction at the 2-$\sigma$ level. With a purple dashed line, we also give the upper bounds on the GW energy density from CMB observations, as explained in App.~\ref{app:CMB}. Dashed black lines illustrate the PBH mass range compatible with the OGLE microlensing events, which could be explained by planet-mass PBHs. We provide a more detailed plot of the typical PBH mass in terms of $f_{\rm p}$ and $w$ in Fig.~\ref{fig:peak mass}. With a red and an orange star, we respectively point the best fit values before and after, including the bound from no PBH overproduction.

\begin{table}
\begin{tabular}{|l|l|l|l|}
\hline
$c_s^2=w$ & $w$                  & $\log_{10}{\cal A}$    & $\log_{10}f_{\rm p}$       \\ \hline
Prior     & Uniform {[}0.01,1{]} & Uniform {[}-5,3{]} & Uniform {[}-8.5,-5{]} \\ \hline
95\%-HDPI & $[0.134,0.867]$      & $[-2.64,1.49]$    & $[-6.99,\text{NaN}]$         \\ \hline
\end{tabular}
\caption{Priors and 95\% HPDI of the posterior marginalized distributions of the Dirac delta case \eqref{eq:diracdelta}.}
\label{tab:Dirac Delta Uncertainty}
\end{table}

From Fig.~\ref{fig:bayesiandelta}, we also see the degeneracies discussed in \S~\ref{subsec:diracdelta}, Eqs.~\eqref{eq:IRdegeneracy} and \eqref{eq:bumpdegeneracy}. Let us start with the ${\cal A}$-$w$ plane. There, we see how increasing the value of $b$ (or, equivalently, lowering the value of $w$) increases the negative power in the ratio $f_{\rm p}/f_{\rm rh}$ and leads to a bigger value of ${\cal A}$. For low enough $w$ ($w<1/3$), we also see that there is a spread in the values of ${\cal A}$ that comes from the factor $c_s^{-4}=w^{-2}$, such that lowering $w$ also allows for smaller ${\cal A}$. Next, in the $f_{\rm p}$-${\cal A}$ plane, we see a degeneracy along the $\mathcal{A}_\Rcal \propto f_{\rm p}^{1+b-|b|}$ direction. Increasing $f_{\rm p}$ requires larger values of ${\cal A}$. Note that because of the term $|b|$ in the exponent, there are two degeneracy lines, one for $b<0$ and one for $b>0$. We do not see them in Fig.~\ref{fig:bayesiandelta} because the range of $w$ is all connected. However, we will see the two degeneracy lines in the log-normal case. Lastly, we see no particular dependence of $f_{\rm p}$ on $w$. This is expected as the $f_{\rm p}$ and $w$ directions are somewhat unrelated.

We also find interesting implications from our analysis. First, from the IR tail, that is $\Omega_\GW \propto f^{2-2|b|}$, we may constrain $w$. In fact, from the $2\sigma$ bounds, we see that $0.13<w<0.87$. This is consistent with analytical scaling arguments, as also pointed out in Ref.~\cite{Inomata:2023zup}. Since the slope of the power law fitting for NANOGrav roughly satisfies $5/2>n_{\mathrm{PTA}}>1$, it yields $|b|<1/2$, or equivalently $1/9<w<1$. We note that, although our results are also qualitatively consistent with the results of Refs.~\cite{Zhao:2023joc,Liu:2023pau}, we find it difficult to compare with their results due to some simplifications in these works. Ref.~\cite{Liu:2023pau} also includes $f_{\rh}$ as a free parameter.

The results become more restrictive once we include information from the PBH counterpart and CMB constraints. However, it is important to be aware that the PBH abundance is exponentially sensitive to several factors, such as the precise formulation one uses, how one computes the threshold or the amount of non-Gaussianity (for the latter see Refs.~\cite{Franciolini:2023pbf, Wang:2023ost, Figueroa:2023zhu, Choudhury:2023fwk,Liu:2023ymk,Li:2023qua}). Thus, any implication from the PBH counterpart should be taken with a grain of salt, and it may differ from the analysis of Refs.~\cite{Zhao:2023joc,Liu:2023pau,Harigaya:2023pmw} by some ${\cal O}(1-10)$. Here, we use the formalism developed in Refs.~\cite{Kitajima:2021fpq} based on peak theory, which we believe is the currently most accurate method. Including the PBH counterpart, we first find that no PBH overproduction excludes $w=1/3$ at $2\sigma$. To be more precise, we find $w\gtrsim0.42$. We also note that for $w>0.8$ there is no PBH overproduction for the 2-$\sigma$ contour. This is consistent with \cite{Balaji:2023ehk,Zhao:2023joc,Liu:2023pau,Harigaya:2023pmw} where increasing the equation of state (or the sound speed) suppresses PBH formation.

Including bounds from CMB, the parameter space is further bounded. We can zoom in on the $f_{\rm p}$-$w$ plane in the Bayesian contour plot to discuss the allowed PBH mass considering constraints imposed by our CMB and PBH results in Fig. \ref{fig:peak mass}. We conclude that the available parameter space compatible with PTAs, without PBH overproduction, and obeys the CMB constraint is roughly given by $0.42\lesssim w\lesssim 0.87$ and predicts a typical PBH mass range between $2.5\times10^{-5} M_\odot \lesssim  M_{\rm PBH} \lesssim  0.09 M_\odot$. If the planet-mass PBHs can further explain OGLE microlensing events, the overlapped parameter space gives $0.42\lesssim w \lesssim 0.50$. It is remarkable that by combining PTAs, CMB, and PBHs, one can considerably narrow down the value of $w$. Although our conclusions align with the results of Ref.~\cite{Inomata:2023zup}, it is worth noting that because Press-Schechter significantly underestimates PBH production, we have excluded the case of $w=1/3$ in this paper at a 2-$\sigma$ level. Interestingly, the upper bound of $\mathcal{O}(0.1)M_\odot$ could be related to candidate merger events in the LVK data with sub-solar mass compact objects \cite{Clesse:2020ghq,Morras:2023jvb,LIGOScientific:2022hai,Carr:2023tpt,Prunier:2023cyv}.

\subsection{Results for Finite width peak \label{subsec:resultsfinite}}

\begin{figure}
\includegraphics[width=0.6\linewidth]{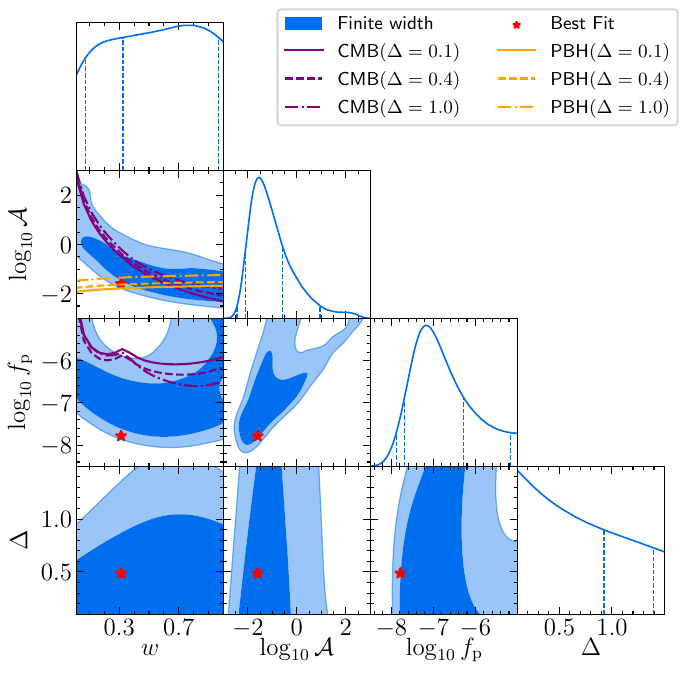}
    \caption{Posterior distribution inferred from the NANOGrav spectrum probability distribution function for the log-normal delta primordial spectrum \eqref{eq:log-normal perturbation}. In orange and purple, we respectively show the constraints from no PBH overproduction (see App.~\ref{sec: PBH}) and the effective number of relativistic species in the CMB (see App.~\ref{app:CMB}). The solid, dashed, and dotted-dashed lines, respectively, denote the constraints for $\Delta=0.1$, $\Delta=0.4$, and $\Delta=1$. For the PBH constraints,  we show the bound calculated by the Press-Schechter method developed in \cite{Domenech:2020ers} (orange curves). The red stars show the best fit parameters. The blue dashed lines in the marginalized distributions show the 68\% and 95\% HDPI in the marginalized distributions.
    }
    \label{fig:bayesianbroad}
\end{figure}

We now include a finite width $\Delta$ in the primordial spectrum using the log-normal peak \eqref{eq:log-normal perturbation}. We use the code \textsc{SIGWfast} to obtain more accurate induced GW spectra for different widths, especially for scales around the peak. This is an important feature since a broader primordial spectrum leads to a broader peak in the induced GW spectrum, which may also fit the PTA data as well as the infrared tail. This was not possible for the Dirac delta case as the resonant peak in the induced GW spectrum is very sharp. Our results for the prior and the 95\% HPDI of the marginalized distributions are displayed in Tab.~\ref{tab:Finite Delta Uncertainty}. In Fig.~\ref{fig:bayesianbroad}, we present the posterior distribution with the NANOGrav dataset, together with the constraints from PBH (orange curves) and CMB (purple curve), respectively. The solid, dashed, and dotted-dashed lines show the constraints from $f_\mathrm{PBH}=1$ for $\Delta=0.1$, $\Delta=0.4$, and $\Delta=1$, respectively, calculated by the simplest Press-Schechter formalism with a Harada-Yoo-Kohri threshold \cite{Harada:2013epa} used in \cite{Domenech:2020ers}. Note that in Fig.~\ref{fig:bayesianbroad}, the constraints are looser than the monochromatic case in Fig.~\ref{fig:bayesiandelta}. This is because when the total power is normalized to be $\mathcal{A_R}$, increasing $\Delta$ will decrease the peak value of the spectrum ($\sim\mathcal{A_R}/(\sqrt{2\pi}\Delta)$), thus the PBHs generated are suppressed significantly. The red star shows the best fit values.

As we include one more parameter $\Delta$, the $1$- and 2-$\sigma$ contours in Fig.~\ref{fig:bayesianbroad} are more complicated than in the Dirac delta case in Fig.~\ref{fig:bayesiandelta}. However, we may understand the main effects of $\Delta$ as follows. On the one hand, since the log-normal peak \eqref{eq:log-normal perturbation} is normalized, the amplitude of the resulting induced GW spectrum around the peak does not change significantly with $\Delta$. Although we see from Eq.~\eqref{eq:IRfinite} that the far IR tail is sensitive to large values of $\Delta$, it still does not matter much with the current data. These are the reasons why in Fig.~\ref{fig:bayesianbroad} the values of ${\cal A}$, $f_{\rm p}$ and $w$ are not sensitive to the precise value of $\Delta$. On the other hand, and most importantly, the presence of $\Delta$ introduces two distinct regimes roughly separated at $f_{\rm p}\sim 10^{-6}\,{\rm Hz}$ where the PTA data is fitted either by $(i)$ the peak in the induced GW spectrum for $f_{\rm p}<10^{-6}\,{\rm Hz}$ or $(ii)$ the IR tail for $f_{\rm p}>10^{-6}\,{\rm Hz}$. This provides different features in the $f_{\rm p}$-$w$ and the $f_{\rm p}$-${\cal A}$ planes with respect to the Dirac delta case, which we explain below.

\begin{table}
\begin{tabular}{|l|l|l|l|l|}
\hline
$c_s^2=w$ & $w$                  & $\log_{10}{\cal A}$  & $\log_{10}f_{\rm p}$  & $\Delta$             \\ \hline
Prior & Uniform {[}0.01,1{]} & Uniform {[}-5,3{]} & Uniform {[}-8.5,-5{]} & Uniform {[}0.1,1.5{]} \\ \hline
95\%-HDPI & $[0.073,0.979]$      & $[-2.43,0.95]$ & $[-7.88, -5.17]$ & $[\text{NaN},1.40]$ \\ \hline
\end{tabular}
\caption{Priors and 95\% HPDI of the posterior marginalized distributions for the log-normal case \eqref{eq:log-normal perturbation}.}
\label{tab:Finite Delta Uncertainty}
\end{table}

First, look at the $f_{\rm p}$-$w$ plane. It is clear that for $f_{\rm p}<10^{-6}\,{\rm Hz}$ there is barely any degeneracy. This is because the shape of the peak of the induced GW spectrum is not very sensitive to $\Delta$ for $\Delta>0.1$. On the other hand, when $f_{\rm p}>10^{-6}\,{\rm Hz}$ the IR tail enters the PTA window. This time, because the primordial spectrum is broad, the IR tail goes as $\Omega_\GW \propto f^{3-2|b|}$. Then, from the power law fitting $5/2>n_{\mathrm{PTA}}>1$ we conclude that there are two disconnected ranges for $b$, namely $-1/4>b>-1/2$ ($5/9<w<1$) and $1>b>1/4$ ($0<w<1/5$). Because of this, the $f_{\rm p}$-$w$ plane has a half doughnut shape. Remarkably, the value of $w=1/3$ is excluded at more than $2\sigma$ for $f_{\rm p}>10^{-6}\,{\rm Hz}$. We also notice that the doughnut shape of the $f_{\rm p}$-$w$ plane is partly carried over to the $f_{\rm p}$-${\cal A}$ plane. From Eq.~\eqref{eq:IRfinite}, the degeneracy in the $f_{\rm p}$-${\cal A}$ is now roughly given by $\Acal_{\cal R}\propto f^{3/2+b-|b|}_{\rm p}$. Since for $f_{\rm p}>10^{-6}\,{\rm Hz}$ the allowed range for $b$ is disconnected, we can see the two degeneracy lines. Unfortunately, we also find that such an interesting regime is mostly incompatible with CMB constraints, shown in purple lines in Fig.~\ref{fig:bayesianbroad}.

\section{Conclusion and Discussion}
\label{sec: c&d}

There is tentative evidence of a GW background at nHz frequencies reported by PTAs \cite{EPTA:2023fyk,EPTA:2023sfo,EPTA:2023xxk,Zic:2023gta,Reardon:2023gzh,Reardon:2023zen,NANOGrav:2023hde,NANOGrav:2023gor,InternationalPulsarTimingArray:2023mzf,Xu:2023wog}, which could be explained by GWs induced by primordial fluctuations. Since these GWs were generated in a period of the early universe that has not been directly probed by other means, we included the EoS $w$ of the early universe as a free parameter in our analysis. For the primordial spectrum of fluctuations, we considered a Dirac delta and a log-normal peak (with a free logarithmic width $\Delta$), inspired by several models of inflation \cite{Kinney:2005vj, Martin:2012pe, Cheng:2018qof, Byrnes:2018txb, Pi:2022zxs, Cheng:2023ikq,Cai:2018tuh, Chen:2019zza,Kawasaki:1997ju, Kawasaki:2012wr, Pi:2017gih}. We then analyzed the NANOGrav data set with \textsc{PTArcade} \cite{Mitridate:2023oar} for the Bayesian analysis and \textsc{SIGWfast} \cite{Witkowski:2022mtg} for the numerical calculations of the induced GWs. Our results for the posterior distributions for the Dirac delta and log-normal can be respectively found in Figs.~\ref{fig:bayesiandelta} and \ref{fig:bayesianbroad}. 

For the Dirac delta spectrum, we found that $0.13\lesssim w\lesssim 0.87$ at $2\sigma$, which is consistent with previous analytical estimations \cite{Inomata:2023zup} (namely $1/9<w<1$). Most interestingly, while standard radiation domination ($w=1/3$) is in the middle of the $1\sigma$ contours using the GW data only, it is excluded at $2\sigma$ when including the upper bounds from no PBH overproduction. Thus, our results on induced GWs and PBHs imply $0.42\lesssim w\lesssim 0.87$ when the universe had temperatures above $100\,{\rm MeV}$. We also investigated the typical mass range of the PBHs, combining the PBH and CMB bounds. We found that, including the $w$ dependence, it is given by $2.5\times10^{-5} M_\odot \lesssim  M_{\rm PBH} \lesssim  0.09 M_\odot$, which can still explain the OGLE microlensing events \cite{Niikura:2019kqi} as well for a narrow range of the EoS, namely $0.42\lesssim w\lesssim 0.5$. Also, the upper bound of $O(0.1)M_\odot$ could explain the candidate merger events in the LVK data with sub-solar mass compact objects \cite{Clesse:2020ghq,Nitz:2022ltl,Morras:2023jvb,LIGOScientific:2022hai,Carr:2023tpt,Prunier:2023cyv}. As this is out of the scope of this paper, we leave a detailed study for future work.
Note that our conclusions slightly differ from Ref.~\cite{Inomata:2023zup} because of the different formalism adopted for estimating the PBH abundance. Nevertheless, based on peak theory other than the simplest Press-Schechter formalism, our analysis gives the most accurate estimate for the PBH bounds.

For the log-normal spectrum, we find that, in general terms, the GW data cannot constrain the EoS $w$. The main reason is that while the Dirac delta case has a precise prediction for the low frequency tail of the induced GWs that enters the PTA window, there are different possibilities for the log-normal depending on the width $\Delta$. For example, for sharp peaks ($\Delta\lesssim {\cal O}(0.1)$) there is the near-IR tail and the far-IR tail which respectively scale as $f^{2-2|b|}$ ($f_p\gg f\gg\Delta \times f_{\rm p}$) and $f^{3-2|b|}$  ($f\ll \Delta \times f_{\rm p}$). For broad peaks ($\Delta\gtrsim {\cal O}(0.1)$), the induced GW is also a broad peak, the shape of which is not very sensitive to the value of $w$. Thus, with the current precision, we cannot constrain the value of $w$. We only find a faint preference for small $\Delta$ in the NANOGrav data set and that the best-fit point fits the PTA data with the peak of the induced GW. To explore this interpretation further, namely the log-normal peak, one would need an experiment with a much larger frequency range. Or, one could also discuss specific models to remove the uncertainty in $\Delta$. Regarding the PBH counterpart, we find that while a stiffer EoS $w$ alleviates the possible PBH overproduction, a broader log-normal peak also reduces the abundance of the PBH counterpart.

In the future, we will be able to combine the likelihood from PTAs, micro-lensing observations as well as ground and space-based detectors such as $\mu$-Ares \cite{Sesana:2019vho}, binary resonances \cite{Blas:2021mpc,Blas:2021mqw} and the Nancy Roman telescope \cite{Wang:2022sxn} (in the $\mu$Hz regime), LISA and Taiji \cite{Barke:2014lsa,Ruan:2018tsw} (for mHz frequencies), TianQin \cite{Gong:2021gvw} and  DECIGO \cite{Yagi:2011wg,Kawamura:2020pcg} (for frequencies around Hz), Einstein Telescope \cite{Maggiore:2019uih}, Cosmic Explorer \cite{ce} and Voyager \cite{A+,voyager} (for GWs with Hz-kHz frequencies). In this way, we will have access to the induced GW spectrum in a much broader frequency range. At the same time, the PBH counterpart and the GW from the mergers of such PBH binaries \cite{Inomata:2023zup} will allow us to do a precise Bayesian analysis to probe primordial fluctuations on small scales as well as the content of the early universe. And, perhaps, confirm (or rule out) the induced GW interpretation of the PTA data.

Our work could also be improved in several ways. First, the calculations of the PBH abundance for a finite-width peak require refinements, mostly due to the high momentum tail of the primordial scalar perturbation and the complexity of the threshold in peak theory. Second, we could include the position of the reheating frequency $f_\rh$ as a free parameter. To do so, one should improve the kernels for $f<f_{\rm rh}$ of Refs.~\cite{Domenech:2019quo,Domenech:2020kqm}. One could also consider a gradual reheating instead of an instantaneous one. Lastly, one may consider including primordial non-Gaussianity of primordial fluctuations.

\acknowledgments
We are grateful for the insightful discussions with Albert Escriv\`{a} and Cristian Joana, especially for the numerical data for the collapse threshold for a Dirac delta primordial spectrum and $w<1/3$ provided by Albert Escriv\`{a}, which can be found in Fig.~11 of Ref.~\cite{Escriva:2022duf}. This work is supported in part by the National Key Research and Development Program of China Grant No. 2021YFC2203004. G.D. is supported by the DFG under the Emmy-Noether program, project number 496592360. S.P. is supported by Project No. 12047503 of the National Natural Science Foundation of China, by the JSPS Grant-in-Aid for Early-Career Scientists No. JP20K14461, and by the World Premier International Research Center Initiative (WPI Initiative), MEXT, Japan.

\appendix

\section{The Primordial Black Hole counterpart}
\label{sec: PBH}

In this section, we discuss the PBH formation when the amplitude of the curvature perturbation is strong enough to generate scalar-induced GWs as the SGWB detected by PTAs. We use the peak theory, following Ref.~\cite{Kitajima:2021fpq}, to calculate the PBH formation for the monochromatic spectrum. For the finite-width case (i.e. a lognormal peak in the power spectrum), to avoid the complexity of the calculation and the uncertainty of the threshold in the finite-width case, we simply apply the Press-Schechter method with the Harada-Yoo-Kohri threshold \cite{Harada:2013epa}, following the calculations in Ref.~\cite{Domenech:2020ers}. We draw the $f_\mathrm{PBH}=1$ lines for the monochromatic case and for a few different widths in Fig.~\ref{fig:bayesiandelta} and Fig.~\ref{fig:bayesianbroad} respectively, with a fixed frequency $f_{\rm p}=10^{-7}$ Hz. This is because the dependence of the PBH mass function on the peak frequency $f_{\rm p}$ is relatively weak compared to the amplitude $\Acal_\Rcal$, which will be explained in details in following context.

In the peak theory, whether a PBH will form around a peak of the density fluctuation depends on whether the compaction function exceeds the profile-dependent threshold. We follow the definitions and calculations in Ref.~\cite{Kitajima:2021fpq}, but generate the threshold to a general EoS $w$, discussed in \cite{Escriva:2022duf}. The horizon mass during that period at the horizon reentry of a comoving wavenumber $k$ 
\begin{align}
    M(k)&=M_{\rh} \left(\frac{k}{k_\rh}\right)^{-\frac{3(1+w)}{1+3w}},\\
    M_{\rh}&=5.33\times10^{35}\left(\frac{g_*}{10.75}\right)^{-1/6}\left(\frac{T_\rh}{20\text{MeV}}\right)^{-2} \mathrm{g},
\end{align}
where $M_{\rh}$ stands for the horizon mass at the end of reheating. When the physical radius $R_m=a e^{\hat{\zeta}(r_m)} r_m$ reenter the horizon, $a e^{\hat{\zeta}(r_m)} r_m H = 1$ , the horizon mass
\begin{equation}
    M_H\equiv M(a/R_m)=M(k_{\rm p})\left(x_m e^{\hat{\zeta}_m(\tilde{\mu}_2)}\right)^{\frac{3(1+w)}{1+3w}},
\end{equation}
where $x_m=r_m k_{\rm p}$.

The relation between the PBH mass and the horizon mass follows a scaling relation around the threshold
\begin{equation}
    M\left(\mu_2, \tilde{k}_3,w\right)=\mathcal{K}\left(\tilde{k}_3,w\right)\left(\tilde{\mu}_2-\tilde{\mu}_{2,\text{th}}\left(\tilde{k}_3,w\right)\right)^{\gamma(w)} M_H\left(\tilde{\mu}_2, \tilde{k}_3,w\right)    
\end{equation}
where $\tilde{\mu}_2$ and $\tilde{k}_3$ describe the height and the width of the peak. The probability distribution of   
them is given by the power spectrum of the random field, and $\tilde{k}_3$ is restricted to 1 in monochromatic case \cite{Kitajima:2021fpq}. $\gamma(w)$ can be expressed semi-analytically, given in \cite{Maison:1995cc,Koike:1999eg}, which has been confirmed numerically in \cite{Musco:2012au,Escriva:2022duf}. $\mathcal{K}$ has a $O(1)$ value which depends on shape as well as $w$ and we choose $\mathcal{K}=1$ in our calculations. We only need the threshold for different EoS parameters $w$. For $w>1/3$, the threshold almost only depends upon the compaction function
\begin{equation}
    \hat{\mathcal{C}}(r)=\frac{2}{3}\left[1-\left(1+r \hat{\zeta}^{\prime}\right)^2\right]
\end{equation}
around its peak value, which can be expressed by an analytical formula up to the second order derivative of the compaction function \cite{Escriva:2020tak,Escriva:2022duf}. The numerical results are used for $w<1/3$, which can be found in Fig.~11 of Ref.~\cite{Escriva:2022duf}. Thus, the mass function of PBH can be written as 
\begin{equation}\label{eq: Mass function}
    f_{\mathrm{PBH}}(M)=\left|\frac{d \ln M}{d \mu_2}\right|^{-1}n_\mathrm{pk}(\tilde\mu_2(M))~M\propto {k_{\rm p}}^\frac{6w}{1+3w},\\
\end{equation}
where 
\begin{align}
    \label{peaknum}
    n_\mathrm{pk}(\tilde\mu_2)&=\left(\frac{3}{2\pi}\right)^{3/2}k_{\rm p}^3 f(\tilde{\mu}_2/\sqrt{\Acal_\Rcal})\frac{1}{\sqrt{2\pi \Acal_\Rcal}} \exp\left(-\frac{\tilde{\mu}_2^2}{2\Acal_\Rcal}\right),\\
    \left|\frac{d \ln M}{d \mu_2}\right|&= \left|\frac{3(1+w)}{1+3w}\hat{\zeta}_m'(\mu_2)+\frac{\gamma(w)}{\mu_2-\mu_{2,\mathrm{th}}}\right|,
\end{align}
and $\hat{\zeta}'(\mu_2)\equiv d \hat{\zeta}(x_m)/ d\tilde{\mu_2}$. 
The $f$ function is the number density of peaks and $f(x)\sim x^3$ at high tails when $\tilde\mu_2/\sqrt{\mathcal{A}_\Rcal}\gg1$ \cite{Bardeen:1985tr}.\footnote{The concrete form of $f$ in Eq. \eqref{peaknum} is given by
\begin{align} 
f(\xi)= & \frac{1}{2} \xi\left(\xi^2-3\right)\left(\operatorname{erf}\left[\frac{1}{2} \sqrt{\frac{5}{2}} \xi\right]+\operatorname{erf}\left[\sqrt{\frac{5}{2}} \xi\right]\right)\notag \\ 
& +\sqrt{\frac{2}{5 \pi}}\left\{\left(\frac{8}{5}+\frac{31}{4} \xi^2\right) \exp \left[-\frac{5}{8} \xi^2\right]+\left(-\frac{8}{5}+\frac{1}{2} \xi^2\right) \exp \left[-\frac{5}{2} \xi^2\right]\right\}
\end{align}
}
The total mass factor can be given by integrating over Eq.\eqref{eq: Mass function},
\begin{equation}
    f_{\mathrm{tot}}=\int f_{\mathrm{PBH}}(M) d\ln M.
\end{equation}
The PBH abundance is mainly determined by the $\exp(-\tilde{\mu}^2_2/(2\Acal_\Rcal))$ in Eq. \eqref{eq: Mass function} when $\tilde{\mu}_{2,\mathrm{th}}\gg \sqrt{\Acal_\Rcal}$, which indicates the PBH bound almost has $\Acal_\Rcal(f_\mathrm{PBH}=1)\propto \tilde{\mu}_{2,\mathrm{th}}^2$ with a slight dependence on $f_{\rm p}$ from the $k_{\rm p}^\frac{6w}{(1+3w)}$ term in Eq.\eqref{eq: Mass function}. So we directly fix $f_{\rm p}=10^{-7}$ Hz and plot the corresponding upper bound for the amplitude of the scalar perturbation for different EoM $w$ in Dirac-delta peak case in Fig.~\ref{fig:bayesiandelta}. For simplicity, we use the 
Press-Schechter formalism to give a conservative estimation for the upper bound for the peak with different widths, shown in Fig.~\ref{fig:bayesianbroad}. Note that the PBH constraints are with respect to $\Acal_\Rcal$, but as we pointed out earlier, this at most contributes a factor $\sim$ 2.

After obtaining the mass function by Eq. \eqref{eq: Mass function}, we can check whether the planet-mass PBH can explain the ultrashort-timescale microlensing events observed by OGLE, firstly proposed in \cite{Domenech:2020ers}. In \cite{Niikura:2019kqi}, the PBH mass and abundance needed to explain these events are given at 95\% C.L. for the monochromatic mass function, which is about 1\% to 10\% of the dark matter at around $10^{-4}$ to $10^{-6}~M_\odot$. Reducing the PBH abundance from $f_\mathrm{PBH}=1$ to $f_\mathrm{PBH}=0.01$ only reduces the amplitude of the perturbation very slightly, so we directly substitute the upper bound amplitude for $f_\mathrm{PBH}=1$ to calculate peak mass for different parameters, which is displayed in Fig.~\ref{fig:peak mass}. It is a zoom-in of the $f_{\rm p}$-$w$ subfigure in Fig.~\ref{fig:bayesiandelta}. We can see that there is still an allowed region in the parameter space, which means that the OGLE ultrashort-timescale microlensing events and the nHz SGWB can be simultaneously explained. 

\section{Constraints from CMB Anisotropy\label{app:CMB}}

The GW background, acting as an additional radiation component, may affect Big Bang Nucleosynthesis (BBN) and the anisotropy of the CMB. The additional energy density in the form of radiation is conventionally parameterized by $N_{\mathrm{eff}}$, the effective number of neutrino species after electron-positron annihilation. One may translate the effective number of relativistic species $\Delta N_{\mathrm{eff}}$ to an upper bound on the integrated energy density of GW background, namely \cite{Binetruy:2012ze}
\begin{equation}\label{eq:CMBconstraint}
    \Omega_{\rm GW,0}h^2 \leq 5.6 \times 10^{-6}\left(\frac{\Omega_{r,0}h^2}{2.47 \times 10^{-5}}\right)\left(\frac{g_S\left(T_0\right)}{3.91}\right)^{4 / 3}\left(\frac{g_S\left(T_0\right)}{3.91}\right)^{-4 / 3} \Delta N_{\mathrm{eff}}.
\end{equation}
Since BBN and CMB yield similar results, we choose the constraints from the CMB, $\Delta N_{\mathrm{eff}} < 0.30$ (i.e $(h^2 \rho_{\mathrm{GW}}/\rho_c)_0 \leq 1.7 \times 10^{-6}$) at 95 \% \cite{Planck:2018vyg}, as our conservative estimation.

We compute the total energy density from the induced GWs as follows. First, we parametrize the primordial spectrum as $\Pcal_{\Rcal,k_{\rm p}}(k)=\Acal_\Rcal F(k/k_{\rm p})$ (which could be the Dirac delta \eqref{eq:diracdelta} or the log-normal peak \eqref{eq:log-normal perturbation}). Then, we integrate the induced GW spectrum over all frequencies as
\begin{align}\label{eq:Omegatot}
    \Omega^{\mathrm{tot}}_{\GW,\rh}&= \Acal^2 \left(\frac{f_{\rm p}}{f_{\rh,*}}\right)^{-2b} \int_0^{\infty} d \ln\frac{f}{f_{\rm p}}\, Q_w\left(\frac f{f_{\rm p}}\right)\,,
\end{align}
where we defined
\begin{align}
Q_w(x)&= x^{-2b}\int_0^1 \mathrm{~d} d \int_1^{\infty} \mathrm{d} s\, \mathcal{T}_w(d, s)\, F\left(x(s+d)\right) F\left(x(s-d)\right).
\end{align}

In the Dirac delta case, Eq.~\eqref{eq:Omegatot} has four free parameters: $f_{\rm p}$, $w$ and ${\cal A}$. In the log-normal case, we have an additional parameter $\Delta$. Then, for any given $w$ and $\Delta$, we find the relation ${\cal A}(f_{\rm p})$ such that the induced GW spectrum goes through the geometric mean of the energy spectral density with the highest probability in different frequency bins, called $\Omega_{\GW,\rm PTA}$. This provides a fixed point in the $(f,\Omega_{\GW})$ domain. Further assuming that $f_{\rm p}\gg f_{\rm PTA}$, \textit{i.e.} the peak frequency is much larger than the PTA frequencies, we may use the IR tail of the induced GW spectrum to find the relation ${\cal A}(f_p)$. Explicitly, we substitute the IR limit of Eq.~\eqref{eq:omegarh} into Eq.~\eqref{eq:omegatoday}, 
\begin{align}\label{eq:aroffp}
\Omega_{\GW,\rm PTA}=\Omega_{\GW,\rh}(f\ll f_{\rm p})\Big|_{f=f_{\rm PTA}}\approx{\cal A}_{\cal R}^2
\left\{
        \begin{aligned}
            &C^{\delta}(b) \frac{f_{\rm p}}{f_{\PTA}}^{2(1-|b|+b)}     & (\text{Dirac-Delta})\\
            &C^{\Delta}(b) \frac{f_{\rm p}}{f_{\PTA}}^{2(3/2-|b|+b)}   & (\text{Log-normal})
        \end{aligned}
        \right.\,,
\end{align}
where the coefficients $C^{\delta}(b)$ and $C^{\Delta}(b)$ can be calculated numerically (although one could also use the analytical formulas given in \eqref{eq:IRDIRACDELTA} and \eqref{eq:IRfinite}). Eq.~\eqref{eq:aroffp} should be understood as a relation ${\cal A}(f_{\rm p})$ when fixing ($f_{\rm PTA}$, $\Omega_{\GW,\rm PTA}$, $w$, $\Delta$). With such relations, we compute Eq.~\eqref{eq:Omegatot} and impose the CMB bound \eqref{eq:CMBconstraint}. The results are shown in purple lines Figs.~\ref{fig:bayesiandelta} and \ref{fig:bayesianbroad}.

\bibliography{refgwscalar_v2}
\bibliographystyle{apsrev4-1}

\end{document}